\documentclass[preprint,pra,aps,superscriptaddress,showpacs]{revtex4}
\usepackage{epsfig}
\usepackage{graphicx}
\usepackage{amsmath}

\newcommand{\beq}{\begin{eqnarray}}
\newcommand{\eeq}{\end{eqnarray}}
\def\be{\begin{equation}}
\def\ee{\end{equation}}
\def\ba{\begin{eqnarray}}
\def\ea{\end{eqnarray}}

\begin{document}
\title
{Nonlinear Landau-Zener Processes in a Periodic Driving Field}
\author{Qi Zhang}
\affiliation{Department of Physics and Center for Computational
Science and Engineering, National University of Singapore,117542,
Republic of Singapore}
\author{Peter H\"anggi}
\affiliation{Department of Physics and Center for Computational
Science and Engineering, National University of Singapore,117542,
Republic of Singapore} \affiliation{Institut f\"ur Physik,
Universit\"at Augsburg, Universit\"atsstra{\ss}e 1, D-86135
Augsburg, Germany}
\author{Jiangbin Gong}
\email{phygj@nus.edu.sg} \affiliation{Department of Physics and
Center for Computational Science and Engineering, National
University of Singapore,117542, Republic of Singapore}
\affiliation{NUS Graduate School for Integrative Sciences and
Engineering, Singapore
 117597, Republic of Singapore}
\date{\today}
\begin{abstract}
Effects of a periodic driving field on Landau-Zener processes are
studied using a nonlinear two-mode model that describes the
mean-field dynamics of a many-body system. A variety of different
dynamical phenomena in different parameter regimes of the driving
field are discussed and analyzed. These include shifted, weakened,
or enhanced phase dependence of nonlinear Landau-Zener processes,
nonlinearity-enhanced population transfer in the adiabatic limit,
and Hamiltonian chaos on the mean field level. The emphasis of this
work is placed on how the impact of a periodic driving field on
Landau-Zener processes with self-interaction differs from those
without self-interaction. Aside from gaining understandings of
driven nonlinear Landau-Zener processes, our findings can be used to
gauge the strength of nonlinearity and for efficient manipulation of
the mean-field dynamics of many-body systems.

\end{abstract}
\maketitle

\section {Introduction}
Landau-Zener (LZ) transition \cite{Landau} forms a fundamental
dynamical process relevant to a variety of contexts, such as quantum
control and quantum information \cite{LZ1,LZ12,LZ13,LZ4,LZ5},
quantum dots \cite{LZ2}, molecular clusters \cite{LZ3}, to name a
few. If the external bias involved in a LZ process is tuned
adiabatically, then the LZ tunneling probability goes to zero,
yielding a robust scenario for realizing a complete quantum
population inversion. If the external bias is varied
non-adiabatically, then the LZ tunneling offers a diagnosis tool for
understanding the quantum dynamics. Indeed, the paramount importance
of LZ processes has motivated a large body of theoretical and
experimental work \cite{Hanggirev,saitorev}.

As a very recent development, Wubs {\it et al.} \cite{Hanggi}
studied numerically and analytically how a LZ process might be
further manipulated by considering a periodic driving field.  Within
the rotating-wave-approximation (RWA), Wubs {\it et al.}
\cite{Hanggi} showed that a LZ process in the presence of a driving
field possesses two intriguing features. First, the driving field
can induce interesting quantum interference effects between two
well-separated LZ sub-processes, with the final quantum transition
probability dependent upon the phase of the driving field. Second,
contrary to a standard LZ problem, the population inversion
probability approaches zero (instead of unity) if the bias is tuned
adiabatically. These findings may find novel applications of LZ
processes, especially in identifying an unknown phase of a driving
field or in studying decoherence time scales of a system coupled
with an environment.

Generalizing LZ processes for single quantum systems to those
associated with the mean-field dynamics of interacting many-body
systems, one obtains nonlinear LZ (NLZ) processes. Due to their
implications for quantum control of the dynamics of Bose-Einstein
condensates (BEC) \cite{holthaus}, NLZ processes have also attracted
considerable interests. Note that the mean-field dynamics of an
interacting many-body system is necessarily nonlinear, and as a
result simple physical intuitions based on linear LZ processes may
become invalid in nonlinear cases. For example, the adiabatic
following of a quantum state with an external bias may break down in
an NLZ process, even when the external bias varies at an infinitely
slow rate \cite{Liu,Wu,Korsch}.

Extending the work by Wubs {\it et al}. \cite{Hanggi} from the
linear regime to the nonlinear regime, our interest here is in how a
periodic driving field might affect NLZ processes.  It is hoped that
a complete picture of driven NLZ processes will help make the best
use of a driving field in manipulating the mean-field dynamics of
many-body systems. To reach that long-term goal, it is necessary to
first examine how the impact of a periodic driving field on NLZ
processes differs from that on conventional LZ processes.

In our previous work \cite{ZHG},  NLZ processes subject to a
high-frequency and large-amplitude driving field were studied.
There, using a high-frequency approximation, we showed that the
driven NLZ dynamics can be effectively described by a stationary
Hamiltonian. The spectrum of the effective Hamiltonian can have new
degenerate eigenstates and hence display new topological structures
that are absent in the non-driven cases.  The results offer a simple
approach to the complete suppression of NLZ tunneling, even when the
tunneling is doomed to happen in the absence of a driving field.

This work continues to study driven NLZ dynamics, but with the
parameter regimes different from that in Ref. \cite{ZHG}. We place
our emphasis on comparisons between driven (linear) LZ processes and
driven NLZ processes, thus shedding more light on the nonlinear and
mean-field nature of NLZ dynamics. For example, we shall show that
nonlinearity can induce significant quantum population transfer that
is completely absent in linear cases. We will also show that the
extent of population transfer in driven NLZ processes can be, as
compared with driven LZ processes, either much more or much less
sensitive to the phase of the driving field. It is hoped that our
detailed results below will motivate more efforts towards better
understanding and better control of many-body systems.

This paper is organized as follows. In Sec. II we briefly discuss
our two-mode model of NLZ processes in a periodic driving field.  In
Sec. III we discuss the high-frequency approximation in treating NLZ
processes in a high-frequency and large-amplitude driving field. In
Sec. IV we consider a different regime of field parameters, where
the RWA can
be applied. 
In Sec. V  we study another parameter regime, where neither the
high-frequency approximation nor the RWA is valid. Section V
concludes this work.

\section{Model of Nonlinear Landau-Zener processes in an ``off-diagonal" driving field}
Motivated by the model of linear driven LZ processes considered in
Ref. \cite{Hanggi}, we consider a two-mode model of driven NLZ
processes as follows:
\begin{equation}\label{Hamiltonian}
{H}(t)=\frac{1}{2}\left(\begin{array}{cc}\gamma+c(|b|^{2}-|a|^{2})&\Delta_{C}+\Delta_{0}\cos(\omega t+\beta)\\
\Delta_{C}+\Delta_{0}\cos(\omega
t+\beta)&-\gamma-c(|b|^{2}-|a|^{2})\end{array}\right).
\end{equation}
Here, \begin{eqnarray}
\gamma=\alpha t
\end{eqnarray} denotes a time-dependent bias
between two modes of interest and is being varied at a rate
$\alpha$; $|a|^{2}$ and $|b|^{2}$ represent occupation probabilities
on the two modes, with the normalization condition taken as
$|a|^{2}+|b|^{2}=1$.  The terms containing $c$ characterize the
nonlinear self-interaction of a BEC under the mean-field treatment,
with the value of $c$ proportional to the number of bosons and the
s-wave scattering length \cite{holthaus,Liu,Wu,Korsch}; $\Delta_{C}$
represents the static coupling between the two modes; and
$\Delta_{0}$ is the amplitude of an external driving, with the
frequency $\omega$ and the phase parameter $\beta$.  Unlike previous
models for driven LZ/NLZ  processes
\cite{holthaus,saito08,modulationLiu} or for coherent destruction of
tunneling \cite{CDT1991,Hanggirev} (for a recent beautiful
experimental validation of the latter phenomenon see in Refs.
\cite{devalle,oberthaler}), the driving field here appears only in
the off-diagonal terms of the above Hamiltonian and is hence an
``off-diagonal" driving field. That is, the external field directly
modulates the coupling between the two modes, rather than modulating
their energy bias. The initial state of an NLZ process is always
taken as $a=1$ and $b=0$, and the occupation probability $|b|^{2}$
in the end, denoted by $|b(\infty)|^{2}$, is the final quantum
transition probability (or the final population transferred between
the two modes).  Note also that all the variables here should be
understood as scaled dimensionless variables with $\hbar=1$.
Throughout we use $\Delta_{0}$ to scale. Then, $\Delta_{0}=1$,
$\omega$ is in units of $\Delta_{0}/\hbar$, $\alpha$ is in units of
$\Delta_{0}^2/\hbar$, and $c$ is in units of $\Delta_{0}$.

Within the BEC context, the above Hamiltonian may be experimentally
realized in several ways. For example, one may consider a BEC in a
double-well potential \cite{weiss,ober,double-well,oberprl}, with
the barrier height periodically modulated, or by a BEC in an optical
lattice occupying two bands \cite{two-band}, with the well-depth of
the optical lattice periodically modulated. One may also regard the
two modes as two internal states of a BEC, such as $^{87}Rb$
\cite{Rb}, with the energy bias $\gamma$ effectively realized by the
detuning of a coupling field from the resonance and the off-diagonal
modulation achieved by modulating the intensity of the coupling
field. Though we only refer to the BEC context below, it should be
noted that this two-mode model of NLZ might be relevant to physics
of Josephson junctions as well as nonlinear optics \cite{luo}.

In the special case of $\Delta_{0}=0$, the above Hamiltonian reduces
to the well-known model of non-driven NLZ processes \cite{Wu}.
Therein the eigen-spectrum diagram as a function of $\gamma$ is
known to display a loop structure at the tip of the lower (upper)
level for $c>\Delta_{C}$ ($c<-\Delta_{C}$). Such a loop structure,
absent in linear systems, directly leads to a nonzero LZ transition
probability even when $\gamma$ changes adiabatically. This makes it
interesting to examine what happens if an NLZ process is subject to
a periodic driving field as introduced above.

Without loss of generality we will restrict ourselves to the $c>0$
cases (cases with $c<0$ can be mapped to those with $c>0$).  This
$c>0$ assumption requires, for example, an attractive interaction
for bosons in a double-well potential or a repulsive interaction for
bosons in two energy bands of an optical lattice.
 In current non-modulated
double-well BEC experiments, typical values of $|c/\Delta_{C}|$ for
$10^{3}$ bosons range from $\sim 10^{0}$ to $\sim 10^{1}$
\cite{holthaus,oberprl}. For the latter two-band realization of our
two-mode model, the value of $c$ (scaled by $\Delta_{0}$) can be
easily tuned by controlling the maximal depth of an optical lattice
potential \cite{Wu}.

To examine the driven NLZ dynamics (i.e., $\Delta_{0}\ne 0$), we
consider below three different parameter regimes. In the first
high-frequency and large-amplitude regime , a high-frequency
approximation can be used to obtain an effective Hamiltonian. The
second regime is the RWA regime, where the driven dynamics is again
understood in terms of some effective Hamiltonians. In the third
regime, both the driving frequency $\omega$ and the amplitude
$\Delta_{0}$ are comparable to, or smaller than, the nonlinear
parameter $c$. Our task below is to examine the driven NLZ dynamics
in these three regimes.

\section {High-frequency and large-amplitude driving field}
The NLZ dynamics under the condition $\omega,\Delta_{0}\gg
\gamma_0,c,\Delta_{C},\gamma_{0}$ (where $\gamma_0$ is the initial
value of $\gamma$) was first studied by us in Ref. \cite{ZHG} using
a high-frequency approximation. For completeness and for the sake of
comparison with other regimes, here we briefly revisit this regime.
To that end we first introduce another pair of wave function
parameters $(a',b')$
\begin{eqnarray} \label{transformation}
\nonumber a'=\frac{a+b}{2}e^{i\frac{\Delta_{0}}{2\omega}\sin(\omega
t+\beta)}+\frac{a-b}{2}e^{-i\frac{\Delta_{0}}{2\omega}\sin(\omega t+\beta)},\\
b'=\frac{a+b}{2}e^{i\frac{\Delta_{0}}{2\omega}\sin(\omega
t+\beta)}-\frac{a-b}{2}e^{-i\frac{\Delta_{0}}{2\omega}\sin(\omega
t+\beta)}.
\end{eqnarray}
Substituting these two relations into the Hamiltonian in Eq.
(\ref{Hamiltonian}), one finds the equations of motion for
$(a',b')$:
\begin{eqnarray}\label{NewHamiltonian}
\nonumber i\frac{da'}{dt}&=&
\frac{1}{2}[\gamma\cdot \cos(\theta)+c\cdot \cos^{2}(\theta)(|b'|^{2}-|a'|^{2})\\
\nonumber&&+ic\cdot \sin(\theta)\cos(\theta)(a'^{*}b'-a'b'^{*})]a'\\
\nonumber&&+\frac{1}{2}[\Delta_{C}-i\gamma \cdot \sin(\theta)+c\cdot \sin^{2}(\theta)(a'^{*}b'\\
\nonumber&&-a'b'^{*})+ic\cdot
\sin(\theta)\cos(\theta)(|a'|^{2}-|b'|^{2})]b'; \nonumber
\\
 i\frac{db'}{dt}&=&\frac{1}{2}[\Delta_{C}+i\gamma \cdot
\sin(\theta)-c\cdot \sin^{2}(\theta)(a'^{*}b'\nonumber \\
\nonumber&&-a'b'^{*})-ic\cdot \sin(\theta)\cos(\theta)(|a'|^{2}-|b'|^{2})]a'\\
\nonumber&&+\frac{1}{2}[-\gamma \cdot
\cos(\theta)-c\cdot \cos^{2}(\theta)(|b'|^{2}-|a'|^{2})\\
&&-ic\cdot \sin(\theta)\cos(\theta)(a'^{*}b'-a'b'^{*})]b',
\end{eqnarray}
where $\theta\equiv \frac{\Delta_{0}}{\omega}\sin(\omega t+\beta)$.

\begin{figure}[t]
\begin{center}
\vspace*{-0.5cm}
\par
\resizebox *{9cm}{8cm}{\includegraphics*{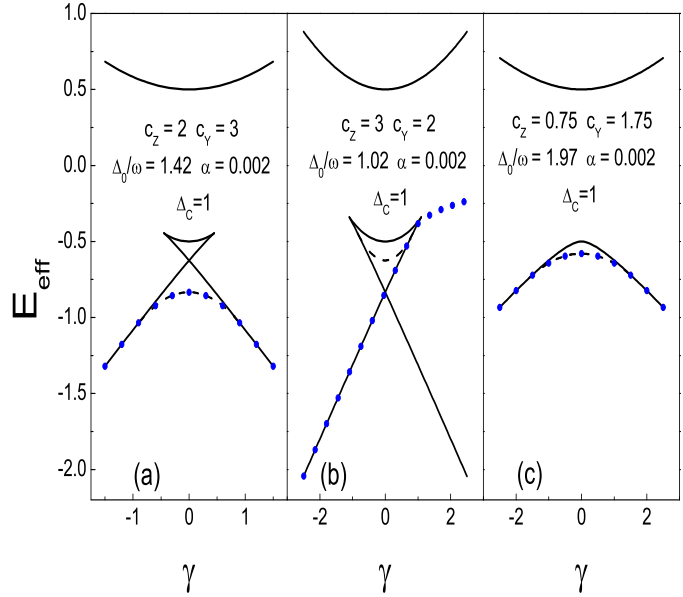}}
\end{center}
\par
\vspace*{-0.5cm} \caption{Eigen-energy structures of the effective
Hamiltonian in Eq. (\ref{effective}) that describes NLZ processes in
a high-frenquency and large-amplitude driving field.  The effective
Hamiltonian is obtained with a high-frequency approximation. The
discrete points represent the time-evolving expectation value of the
mean-field energy when $\gamma$ increases in one adiabatic process.
Along with other parameters, the values of the nonlinear parameter
$c$ and the driving frequency $\omega$ are indicated on the three
panels. As mentioned in the text, all variables are appropriately
scaled and hence dimensionless.} \label{f1}
\end{figure}

For sufficiently large $\omega$, the oscillation in $\theta$ is much
faster than all other time scales of the system. As such, during
each period of $T=2\pi/\omega$, the change in $a'$ and $b'$ is
negligible and considering their averages over $T$ will greatly
reduce the equations of motion. Specifically, upon a time averaging
the odd functions of $t$ in Eq. (\ref{NewHamiltonian}) vanish and
the effective equations of motion become
\begin{eqnarray}\label{Newmotion}
\nonumber i\frac{da'}{dt}&=&\frac{1}{2}[\gamma' +c_{Z}
(|b'|^{2}-|a'|^{2})]a'\\
\nonumber&&+\frac{1}{2}[\Delta_{C}+c_{Y}
(a'^{*}b'-a'b'^{*})]b'\\
\nonumber i\frac{db'}{dt}&=&\frac{1}{2}[\Delta_{C}-c_{Y}(a'^{*}b'-a'b'^{*})]a'\\
&&+\frac{1}{2}[-\gamma'-c_{Z}(|b'|^{2}-|a'|^{2})]b',
\end{eqnarray}
where $\gamma'=\gamma \langle\cos(\theta)\rangle_{T}=\gamma
J_{0}(\Delta_{0}/\omega)$, $c_{Z}=c
\langle\cos^{2}(\theta)\rangle_{T}=c[1+J_{0}(2\Delta_{0}/\omega)]/2$,
$c_{Y}=c\langle\sin^{2}(\theta)\rangle_{T}=c[1-J_{0}(2\Delta_{0}/\omega)]/2$,
and $J_{0}$ is the zeroth order Bessel function of the first kind.
Apparently, these newly defined parameters reflect the action of the
high-frequency driving field.

Based on Eq. (\ref{Newmotion}), we can define the following
effective Hamiltonian:
\begin{equation} \label{effective}
{H}_{\text{eff}}=\frac{1}{2}\left(\begin{array}{cc}\gamma+c_{Z}(|b|^{2}-|a|^{2})&\Delta_{C}+c_{Y}(a^{*}b-ab^{*})\\
\Delta_{C}-c_{Y}(a^{*}b-ab^{*})&-\gamma
-c_{Z}(|b|^{2}-|a|^{2})\end{array}\right),
\end{equation}
where we have replaced $a'$ by $a$, $b'$ by $b$, and so on.
Comparing this effective Hamiltonian with the original one in  Eq.
(\ref{Hamiltonian}) for $\Delta_{0}=0$, it is seen that the
nonlinear parameter $c_{Z}$ can be regarded as a re-scaled parameter
$c$, and the new nonlinear terms containing $c_{Y}$ arise as a
surprise. In addition, the ratio of the two nonlinear parameters
$c_{Z}$ and $c_{Y}$ is given by
$[1+J_{0}(2\Delta_{0}/\omega)]/[1-J_{0}(2\Delta_{0}/\omega)]$, which
is easily adjustable by tuning the ratio $\Delta_{0}/\omega$.

In terms of the mean-field eigen-energies of ${H}_{\text{eff}}$ as a
function $\gamma$, three typical level structures of
${H}_{\text{eff}}$ are depicted in Fig. 1.  The structures
represented by the solid lines in Fig. 1 are also typical in
non-driven NLZ models \cite{Wu}, but the dashed lines (each of them
is associated with double degenerate eigenstates) in Fig. 1 are
entirely induced by the high-frequency driving field.  For an NLZ
process with $\gamma$ increasing at a very small rate of
$\alpha=0.002$ (hence an adiabatic NLZ process), the time-evolving
expectation values of the mean-field energy associated with
${H}_{\text{eff}}$ are shown by the discrete points in Fig. 1.
Clearly,  for the two cases in Fig. 1(a) and 1(c),  the evolution of
the system perfectly follows the new eigenstates (dashed lines). The
driving field can hence dramatically affect the dynamics of
adiabatic following. Indeed, in the case of Fig. 1(a), if there were
no additional levels induced by the driving field, then the loop
structure there will necessarily cause the adiabatic following to
break down (this is analogous to what is observed in non-driven NLZ
models \cite{Wu} and will be also seen below).  For other important
findings in this high-frequency large-amplitude regime, please see
Ref. \cite{ZHG}.

The high-frequency approximation used here is also applicable to
high-frequency and small-amplitude fields. However, if
$\Delta_{0}/\omega<<1$, then $J_{0}(2\Delta_{0}/\omega)\sim 1$,
$c_{Z}\sim c$,  $c_{Y}\sim 0$, and the effective Hamiltonian in Eq.
(6) will essentially reduce to the Hamitonian of Eq. (1) with
$\Delta_{0}= 0$. That is, in cases of $\Delta_{0}/\omega<<1$, the
present high-frequency approximation based on the $(a',b')$
representation does not directly give useful insights into the
dynamics. Fortunately, in these cases the RWA treatment becomes
valid and more advantageous to use. This is discussed in detail in
the next section.

\section {High-frequency and small-amplitude driving field}
\subsection{Rotating-Wave Approximation}
In the high-frequency and
small-amplitude regime, we adopt the RWA to understand the effects
of a driving field on NLZ processes. The off-diagonal coupling term
in Eq. (1) is now understood as a superposition of three terms,
\begin{equation}\label{RWA}
\Delta=\frac{\Delta_{0}}{2}\left[\exp(-i\omega
t-i\beta)+\exp(i\omega t+i\beta)\right]+\Delta_{C},
\end{equation}
where the first two terms represent the circular modulation with
frequency $\omega$ and $-\omega$, and the last constant term can be
regarded as a circular modulation with zero frequency. Consistent
with the high-frequency assumption, the effects of the three
circular modulation terms can be analyzed separately. To illustrate
this, we depict in Fig. 2 a typical numerical result for this
parameter regime. Evidently, the transitions primarily occur in
three small time windows around $\gamma\sim -\omega,0,\omega$, each
of them is associated with one of the three circular modulation
terms in Eq. (\ref{RWA}).

\begin{figure}[t]
\begin{center}
\vspace*{-0.5cm}
\par
\resizebox *{8.4cm}{8.4cm}{\includegraphics*{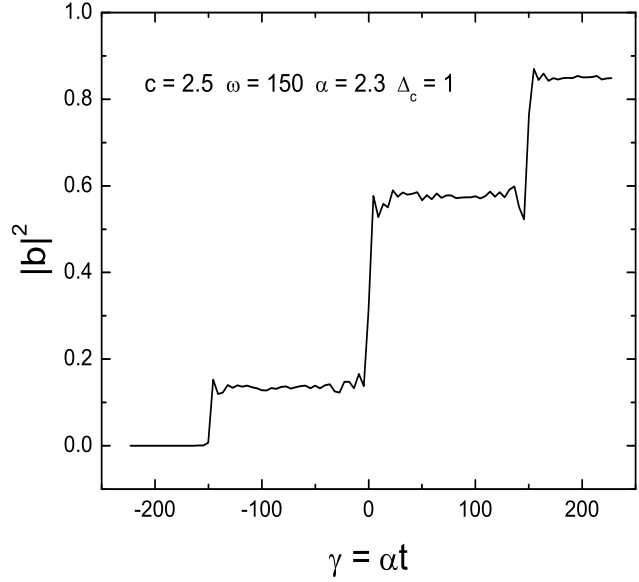}}
\end{center}
\par
\vspace*{-0.5cm} \caption{Transition probability $|b(t)|^{2}$ as a
function of $\alpha t$, with the phase parameter $\beta=0$. The
driving field is in the RWA regime. Three well-separated NLZ
transitions can be seen here.} \label{fig2}
\end{figure}

For the sake of a direct comparison between driven NLZ processes and
the driven LZ processes studied in Ref. \cite{Hanggi}, below we set
$\Delta_{C}=0$. We then only need to consider two circular
modulation terms. The effects of the first circular modulation term
$\frac{\Delta_{0}}{2}\exp(-i\omega t-i\beta)$ become obvious in a
rotating frame. We hence introduce another pair of wavefunction
parameters $(\tilde{a},\tilde{b})$.
\begin{eqnarray}
\nonumber \tilde{a}&=&a\exp\left[-\frac{1}{2}i(\omega t+\beta)\right]\\
\tilde{b}&=&b\exp\left[\frac{1}{2}i(\omega t+\beta)\right].
\end{eqnarray}
Using this representation and neglecting the counter-rotating terms,
one obtains
\begin{equation} \label{motion2}
i\frac{\partial}{\partial t}\left(\begin{array}{c}\tilde{a}\\
\tilde{b}\end{array}\right)=\tilde{H}_{+}(t)\left(\begin{array}{c}\tilde{a}\\
\tilde{b}\end{array}\right),
\end{equation}
where the RWA effective Hamiltonian $\tilde{H}_{+}(t)$ is given by
\begin{equation}\label{fHamiltonian}
\tilde{H}_{+}(t)=\frac{1}{2}\left(\begin{array}{cc}\gamma_{+}+c(|\tilde{b}|^{2}-|\tilde{a}|^{2})&\frac{\Delta_{0}}{2}\\
\frac{\Delta_{0}}{2}&-\gamma_{+}-c(|\tilde{b}|^{2}-|\tilde{a}|^{2})\end{array}\right),
\end{equation}
with $\gamma_{+}=\gamma+\omega$. This Hamiltonian is essentially
identical with that for a non-driven NLZ model, with the bias
parameter $\gamma$ shifted by $\omega$ and with the effective static
coupling term determined by the amplitude of the periodic driving
field.  This explains the observation from Fig. 2 that one main
transition window is roughly at $\gamma\sim-\omega$.  Because
$|b|^{2}=|\tilde{b}|^{2}$, this RWA treatment also indicates that
for a fixed initial state, the quantum population at the end of such
a transition window should not depend on $\beta$. Similar treatments
apply to the other circulation modulation term
$\frac{\Delta_{0}}{2}\exp(i\omega t+i\beta)$, yielding the following
effective Hamiltonian
\begin{equation}\label{f2Hamiltonian}
\tilde{H}_{-}(t)=\frac{1}{2}\left(\begin{array}{cc}\gamma_{-}+c(|\tilde{b}'|^{2}-|\tilde{a}'|^{2})&\frac{\Delta_{0}}{2}\\
\frac{\Delta_{0}}{2}&-\gamma_{-}-c(|\tilde{b}'|^{2}-|\tilde{a}'|^{2})\end{array}\right),
\end{equation}
with $\tilde{a'}=a\exp\left[\frac{1}{2}i(\omega t+\beta)\right]$ and
$ \tilde{b'}=b\exp\left[-\frac{1}{2}i(\omega t+\beta)\right]$. Here
the effective bias is $\gamma_{-}=\gamma-\omega$, and as a result
the main transition window will be close to $\gamma\sim \omega$.

Because the RWA treatment here is seen to be independent of the
nonlinear parameter $c$, it applies as well to those driven (linear)
LZ processes considered in Ref. \cite{Hanggi}.  Hence, for either
NLZ or LZ processes, a periodic driving field is expected to shift
the main quantum transition window at $\gamma\sim 0$ to two new
windows close to $\gamma \sim \mp \omega$ (how the exact locations of
the transition windows depend on $c$ will be discussed later).

\subsection{Phase-dependence weakened by strong nonlinearity}

The above RWA treatment makes it clear that the NLZ dynamics in this
regime can be described by effective Hamiltonians analogous to that
for non-driven cases.  However,  because the final state of the
transition window around $\gamma\sim -\omega$ becomes the initial
state of the second transition window around $\gamma\sim \omega$,
the driven dynamics should be much richer than in non-driven cases.
Indeed, in the linear case with $c=0$, the analytical result from
Ref. \cite{Hanggi} predicts that the final transition probability is
given by

\begin{equation}\label{linearOccupation}
|b(\infty)|^{2}=4\exp(-\pi\Delta_{0}^{2}/8\alpha)\left[1-\exp(-\pi\Delta_{0}^{2}/8\alpha)\right]\cos^{2}\beta,
\end{equation}
where $\beta$ is the phase of the driving field defined above.
Because the transition probability of the first LZ sub-process does
not depend upon $\beta$, the factor $\cos^{2}(\beta)$ in Eq.
(\ref{linearOccupation}) reflects how the net result associated with
the second transition window is affected by the first one.
Alternatively, this $\cos^{2}(\beta)$ factor can be interpreted as a
quantum interference between two LZ sub-processes at two time
windows \cite{Hanggi}.  This intriguing phase dependence might be
useful in determining an unknown value of $\beta$ \cite{Hanggi}.

\begin{figure}[t]
\begin{center}
\vspace*{-0.5cm}
\par
\resizebox *{8.4cm}{8.4cm}{\includegraphics*{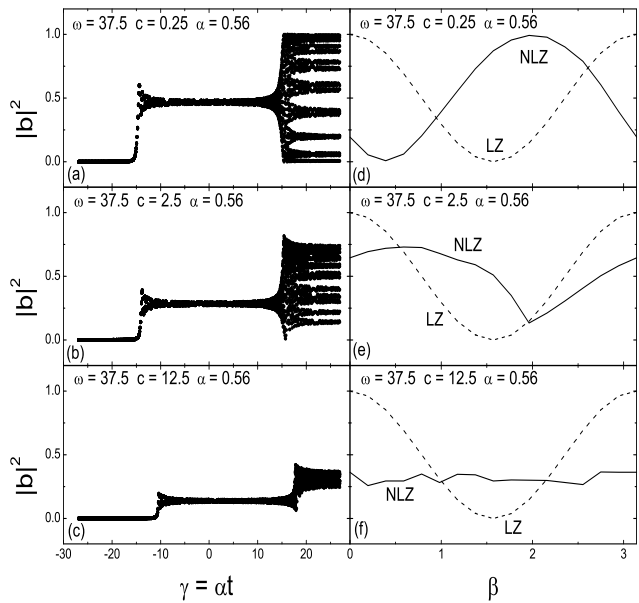}}
\end{center}
\par
\vspace*{-0.5cm} \caption{Left panel: Transition probability
$|b|^{2}$ as a function of $\alpha t$, with $\alpha=0.56$,
$\omega=37.5$, and different values of the nonlinear parameter $c$
and the phase parameter $\beta$.  Right panel: The final NLZ
transition probability $|b(\infty)|^{2}$ (solid lines) as a function
of $\beta$.  Dashed lines represent the theoretical linear result of
Eq. (\ref{linearOccupation}) that displays a $\cos^{2}(\beta)$
dependence.} \label{fig2}
\end{figure}

How does the nonlinear self-interaction of a BEC affect this
$\cos^{2}(\beta)$-dependence of the final transition probability?
Unable to obtain an analytical solution, we choose to examine this
issue computationally. Typical results for a rather rapid ramping of
$\gamma$ are shown in Fig. 3. The ramping rate is chosen to be
$\alpha=0.56$ such that the linear result of Eq.
(\ref{linearOccupation}) yields the maximal transition probability.
Panels (a), (b) and (c) of Fig. 3 depict the time-evolving
occupation probability $|b|^{2}$ as a function of $\alpha t$, with
the initial condition $a=1, b=0$.  In each of the three panels,
results for a number of $\beta$ values are plotted collectively. For
all the shown cases two main quantum transition windows can be
clearly seen, consistent with our expectation based on the RWA
analysis. Comparing Fig. 3(a), 3(b) with Fig. 3(c), it is seen that
as the nonlinear strength $c$ increases, the final transition
probabilities for different values of $\beta$ tend to converge to a
common value, i.e., they become less and less sensitive to $\beta$.
To be more quantitative, we extract the $\beta$-dependence from
Figs. 3(a)-3(c) and then compare it (right panels in Fig. 3) with
the linear result of Eq. (\ref{linearOccupation}).  For the case of
$c=0.25$ (right upper panel of Fig. 3), the $\beta$-dependence of
the final transition probability $|b(\infty)|^{2}$ exhibits a
remarkable phase shift as compared with the linear result (dashed
line). This phase-shift effect might be useful in determining a
small but unknown value of $c$. For cases with sufficiently strong
self-interaction (right bottom panel of Fig. 3), $|b(\infty)|^2$
changes little as $\beta$ varies, constituting a sharp contrast to
the $\cos^{2}(\beta)$ dependence in linear cases.

The weakened $\beta$-dependence due to strong nonlinearity requires
a qualitative explanation. Consider the state after the first
transition window, i.e., the initial state for the second NLZ
sub-process. As $\beta$ varies, the occupation populations
associated with this intermediate state can be hardly changed (as is
evident from the above RWA treatment), but the relative phase
between the amplitudes $\tilde{a'}$ and $\tilde{b'}$ will be
affected. With this understanding, we explain below the observed
weakened $\beta$-dependence of $|b(\infty)|^2$ in terms of the
increasing insensitivity of the second NLZ sub-process on the
quantum phase of its initial state.

Let us map exactly the mean-field dynamics of the effective
Hamiltonian in Eq. (\ref{f2Hamiltonian}) to that of a classical
Hamiltonian system, with the canonical variables
$S=|\tilde{b}'|^{2}-|\tilde{a}'|^{2}$ and $\phi$
\cite{Liu,ZHG,modulationLiu}, with $\phi$ the relative phase between
the amplitudes $\tilde{a}'$ and $\tilde{b}'$.  The mapped classical
Hamiltonian, denoted by $\tilde{H}_{c}$, is then given by \cite{Liu}:
\begin{equation}\label{classical}
\tilde{H}_{c}=\frac{1}{2}\left[-\gamma_{-}
S-\frac{c}{2}S^{2}+\frac{\Delta_{0}}{2}\sqrt{1-S^{2}}\cos\phi\right].
\end{equation}
With this mapping, the phase of the initial state for the second NLZ
sub-process apparently becomes the initial value of the coordinate
$\phi$. Further, the final transition probability $|b(\infty)|^{2}$
is given by $[S(\infty)+1]/2$.

\begin{figure}[t]
\begin{center}
\vspace*{-0.5cm}
\par
\resizebox *{8.4cm}{7cm}{\includegraphics*{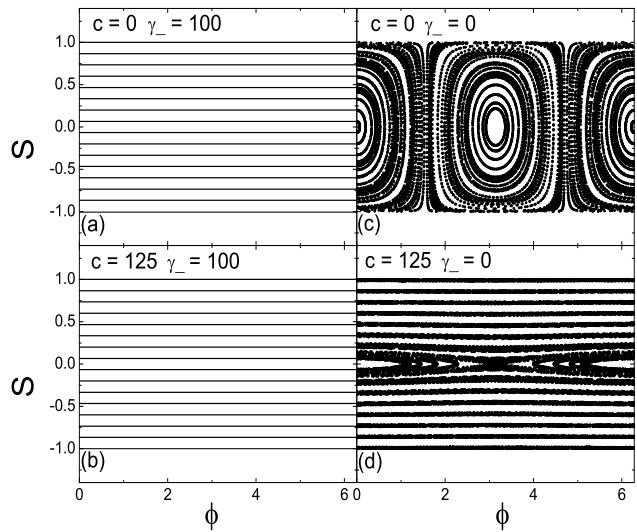}}
\end{center}
\par
\vspace*{-0.5cm} \caption{Phase space structure associated with the
mapped classical Hamtonian in Eq. (\ref{classical}), in terms of the
dimensionless variables $S$ and $\phi$ defined in the text. For
large $c$, most phase space invariant curves will be, independent of
the value of $\gamma_{-}=\gamma-\omega$,  almost straight lines
perpendicular to the $S$-axis.} \label{fig2}
\end{figure}

Figure 4 compares the phase space structure of $\tilde{H}_{c}$ for
$\gamma_{-}=\gamma-\omega=0$ and that for $\gamma_{-}=100$. Consider
first the upper two panels for $c=0$.  In the regime of large
$\gamma_{-}$ [Fig. 4(a)],  the phase space curves become straight
lines perpendicular to the $S$ axis,  with their $S$ values
determining $|b(\infty)|^{2}$. In the regime around $\gamma_{-}=0$
[Fig. 4(c)], the phase space invariant curves rotate around elliptic
fixed points, with their oscillation amplitude in $S$ sensitively
dependent upon $\phi$.  Clearly then, if $\gamma_{-}$ is scanned
rather rapidly, the range of $S$ values for $\gamma_{-}=0$, which
depends on the initial value of $\phi$, will determine the range of
$S$ and hence the possible values of $|b(\infty)|^{2}$. This
qualitatively explains the sensitive $\beta$-dependence in the
linear case. By contrast, as shown in Fig. 4(b) and Fig. 4(d), the
phase space structure for large $c$ is largely independent of
$\gamma_{-}$, with most of the phase space curves being parallel
lines perpendicular to the $S$-axis. The sensitive dependence of
$S(\infty)$ on $\phi$, and hence the sensitive dependence of
$|b(\infty)^{2}|$ on $\beta$, must be weakened for large $c$.
Indeed, in the large $c$ limit, one may only keep the $-cS^2/2$ term
of $\tilde{H}_{c}$. $\tilde{H}_{c}$ then becomes the Hamiltonian for
a ``free-particle" system, with a trivial phase space structure
filled with straight lines perpendicular to the $S$-axis. As such,
the $\beta$-dependence of $|b(\infty)^{2}|$  in the large $c$ limit
will disappear altogether. This completes our qualitative
explanation of the weakened $\beta$-dependence due to increasing
nonlinearity.

\begin{figure}[t]
\begin{center}
\vspace*{-0.5cm}
\par
\resizebox *{8.8cm}{8.8cm}{\includegraphics*{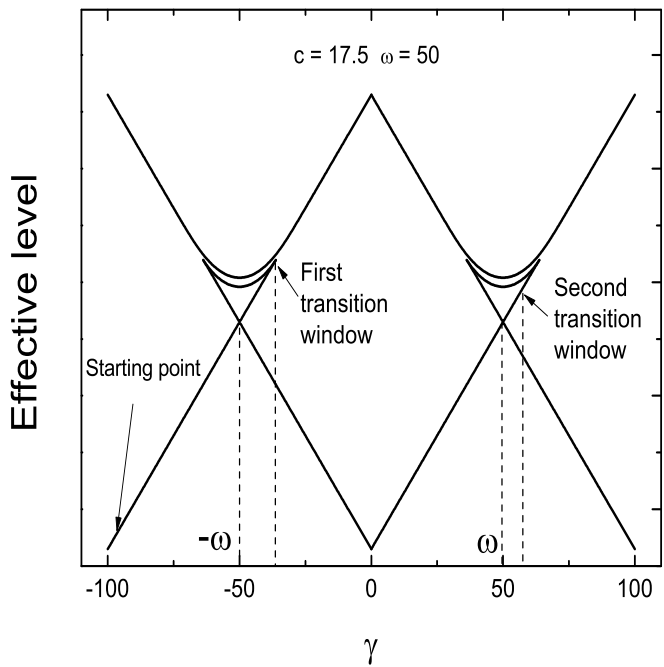}}
\end{center}
\par
\vspace*{-0.5cm} \caption{Schematic energy level structures of the
RWA static Hamiltonians $\tilde{H}_{+}$ in Eq. (\ref{fHamiltonian})
and $\tilde{H}_{-}$ in Eq. (\ref{f2Hamiltonian}).
 $c=17.5$ and
$\omega=50$. All variables are appropriately scaled and hence
dimensionless.}
\end{figure}

Finally, we discuss another interesting observation made from Fig.
3. Comparing Fig. 3(a), 3(b) and 3(c), one notes that the first
transition window is delayed more and more as the nonlinear strength
$c$ increases.  The delay of the second  transition window as a
function of $c$ is however less apparent. To rationalize this
observation we first note that in the cases with negligible $c$, the
standard picture based on linear LZ processes applies and the two
transition windows should be centered precisely at $\gamma\sim
-\omega$ and $\gamma\sim +\omega$. In cases with large $c$, the
effective Hamiltonians $\tilde{H}_{+}$ and $\tilde{H}_{-}$ obtained
in our RWA treatment are expected to display loop structures, in the
same manner as in non-driven NLZ models \cite{Wu}.  One typical
example with $c=17.5$ and $\omega=50$ is shown in Fig. 5, where both
$\tilde{H}_{+}$ and $\tilde{H}_{-}$ display a loop structure in the
lower branch of the energy levels. Due to the loop structure of
$\tilde{H}_{+}$, the increase in $\gamma$ from a very negative value
will not cause any significant transition to the upper branch at
$\gamma \sim  -\omega$. Rather, significant population transfer
occurs only until the system reaches the right ``edge" of the loop,
thus postponing the first transition window. In addition,  because
the loop structure is bigger for larger $c$, the first transition
window will be postponed more as $c$ increases.

Interestingly, after the first NLZ sub-process, the system will in
general occupy both upper and lower levels considerably.  Then, as
$\gamma$ continues to increase in the positive regime, those
populations on the upper branch will make significant transitions at
$\gamma\sim \omega$, whereas those on the lower branch will delay
their transitions again. The overall result is that the precise
location of the second transition window becomes less definitive and
more dependent on the details of the dynamics. With this recognition
we schematically place the second transition window in Fig. 5
between $\gamma\sim\omega$ and the right edge of the right loop
structure.

\subsection{Nonlinearity-induced Transition Probabilities}
Equation (\ref{linearOccupation}) contains another intriguing and
stimulating result for driven linear LZ processes.  That is, in the
adiabatic limit $\alpha\rightarrow 0$, the final transition
probability $|b(\infty)|^{2}$ becomes zero for any $\beta$. The
associated physical picture is as follows.  After the first
transition window of a linear driven LZ process, the adiabatic
population transfer is already complete. But then the system
approaches the second transition window, which transfers all the
population back to the initial mode.  In other words, for linear
driven LZ processes, the transitions associated with the two time
windows $\gamma\sim \mp \omega$ exactly cancel each other in the
adiabatic limit.  Such a strong result for linear systems motivates
us to examine driven NLZ processes in the same adiabatic limit.

\begin{figure}[t]
\begin{center}
\vspace*{-0.5cm}
\par
\resizebox *{8cm}{8cm}{\includegraphics*{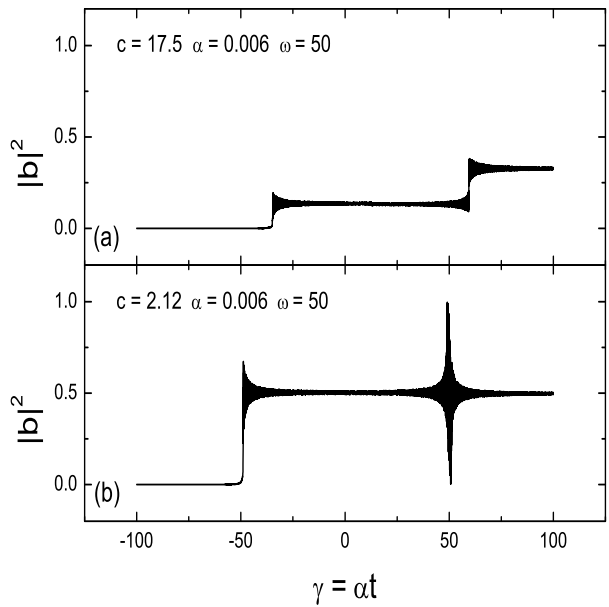}}
\end{center}
\par
\vspace*{-0.5cm} \caption{Transition probability $|b(t)|^{2}$ as a
function of $\alpha t$, for a small value of $\alpha=0.006$ that
simulates the dynamics in the adiabatic limit. Panel (a) and panel
(b) are for two different values of $c$. In each case the results
for a number of $\beta$ values are plotted collectively. The large
final values of $|b|^{2}$ are induced entirely by the nonlinear
nature of the system.} \label{fig2}
\end{figure}

For two values of the nonlinear parameter $c$, Fig. 6 shows the
time-evolving occupation probability $|b|^{2}$, using a sufficiently
small value of $\alpha$ to simulate the dynamics in the adiabatic
limit.  In each case, results with different values of $\beta$ are
plotted collectively. In all these cases, it is seen that the
quantum transition probabilities at the end of the first transition
window are far from unity.  Because in linear cases the adiabatic
population transfer after the first transition window should be
always unity, the incomplete population transfer shown in Fig. 6
after the first transition window constitutes direct evidence of
considerable NLZ tunneling in the adiabatic limit.  Analogous NLZ
tunneling in the second time window can also be expected.  As to the
final transition probability $|b(\infty)|^{2}$ (final values of
$|b|^{2}$), they are seen to be insensitive to $\beta$. This phase
insensitivity is similar to previous non-adiabatic cases with
considerable $c$ values.  But more noteworthy, $|b(\infty)|^{2}$ is
$\sim 0.35$ for $c=17.5$ in Fig. 6(a) and as large as $\sim 0.5$ for
$c=2.12$ in Fig. 6(b). Considering that $|b(\infty)|^{2}$ should be
precisely zero if $c=0$, the large transition probabilities
$|b(\infty)|^{2}$ here are induced entirely by the nonlinear nature
of the system. In addition, comparing Fig. 6(a) and Fig. 6(b), one
also sees that the larger $c$ is, the more delayed the first
transition window will be. This observation is analogous to what is
found in Fig. 3.

To further motivate interests in the nonlinearity-induced transition
probabilities in driven NLZ processes,  we calculate
$|b(\infty)|^{2}$ in many adiabatic cases that cover a wide range of
$c$ values. A typical result, obtained with  $\omega=40$ and
$\alpha=0.003$, is depicted in Fig. 7.  The non-monotonic dependence
of $|b(\infty)|^{2}$ upon $c$ is apparent. In particular, for any
$c<0.5$, the transition probability right after the first transition
window is close to unity and becomes essentially zero after the
second transition window. This behavior is the same as in linear
driven LZ processes in the adiabatic limit.  Hence, for the RWA
regime studied here, the threshold value for observing
nonlinearity-induced population transfer at the end of the adiabatic
process is around $c\sim 0.5$. Figure 7 also indicates that for very
large values of $c$, $|b(\infty)|^{2}$ decays back to zero again.
For intermediate values of $c$, the final transition probability can
be significant. Indeed, the peak value of $|b(\infty)|^{2}$ as a
function of $c$ is found to be as large as $\sim 0.95$.

\begin{figure}[t]
\begin{center}
\vspace*{-0.5cm}
\par
\resizebox *{8.cm}{8.cm}{\includegraphics*{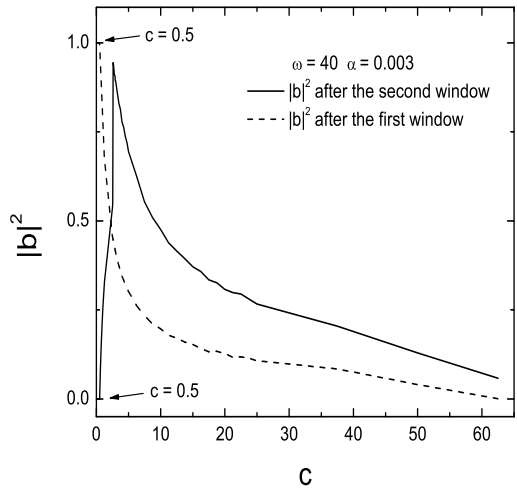}}
\end{center}
\par
\vspace*{-0.5cm} \caption{The final transition probability
$|b(\infty)|^2$ (solid line) as a function of the nonlinear strength
$c$, for a high-frequency and small-amplitude driving field. The
dashed line shows the transition probability right after the first
transition window.}
\end{figure}

Noticing that the $c$ value here is scaled by $\Delta_{0}$, it is
clear that by tuning the amplitude of the driving field one may
realize the peak value of $|b(\infty)|^{2}$ observed in Fig. 7,
without actually tuning the natural scattering length or the number
of bosons of the BEC.  Hence a nearly complete population inversion
is made possible by a periodic driving field in the RWA regime. This
finding offers a potentially useful scenario for improving the
adiabatic population transfer in NLZ processes. Interestingly, in
the high-frequency and large-amplitude regime discussed in Sec. III,
the adiabatic population transfer is improved by suppressing the NLZ
tunneling; whereas the nearly complete population transfer here is
based on the overall effect of two NLZ tunneling processes.

Combining (i) the shifted $\beta$-dependence of $|b(\infty)|^{2}$
for small $c$, (ii) the weakened $\beta$-dependence of
$|b(\infty)|^{2}$ for large $c$, (iii) the delay of the first
transition window as a function of $c$ , and (iv) the
nonlinearity-induced transition probabilities in the adiabatic
limit, one may also design strict tests to check the validity of a
two-mode description of a driven many-body quantum system.

\section {Low-frequency driving fields}
If the driving frequency $\omega$ is comparable to, or smaller than,
$c$ and $\Delta_{0}$, we call it a low-frequency case. In this
regime the previous high-frequency approximation or RWA does not
apply.  As it turns out, complications arise here due to the
emergence of Hamiltonian chaos on the mean-field level.

Using the same approach in Sec. IV-B,  let us first map the driven
NLZ dynamics described by Eq. (1) (with $\Delta_{C}=0$) to that of a
classical Hamiltonian,
\begin{equation}
H_{c}(t)=\frac{1}{2}\left[-\gamma
S-\frac{c}{2}S^{2}+\Delta_{0}\cos(\omega t
+\beta)\sqrt{1-S^{2}}\cos\phi\right],
\end{equation}
where $S$ is given by ($2|b|^2-1)$, and $\phi$ is now the relative
phase between the amplitudes $a$ and $b$.  Because $\omega$ is
small, it is impossible to divide the driven NLZ process into two
sub-processes. But this also implies that, roughly speaking, the
phase space structure at $\gamma\sim 0$ will be most important in
understanding the mean-field dynamics.  As such, we show in Fig. 8
the typical behavior of the Poincar\'{e} surface of section of
$H_{c}(t)$ with $\gamma=0$, $\omega=1$, $\beta=0$, and with the
nonlinearity parameter $c$ ranging from 0 to 250. It is seen that
for intermediate values of $c$, the phase space is completely
chaotic, whereas for small or large values of $c$, the dynamics is
primarily regular.

\begin{figure}[t]
\begin{center}
\vspace*{-0.5cm}
\par
\resizebox *{8.8cm}{8.8cm}{\includegraphics*{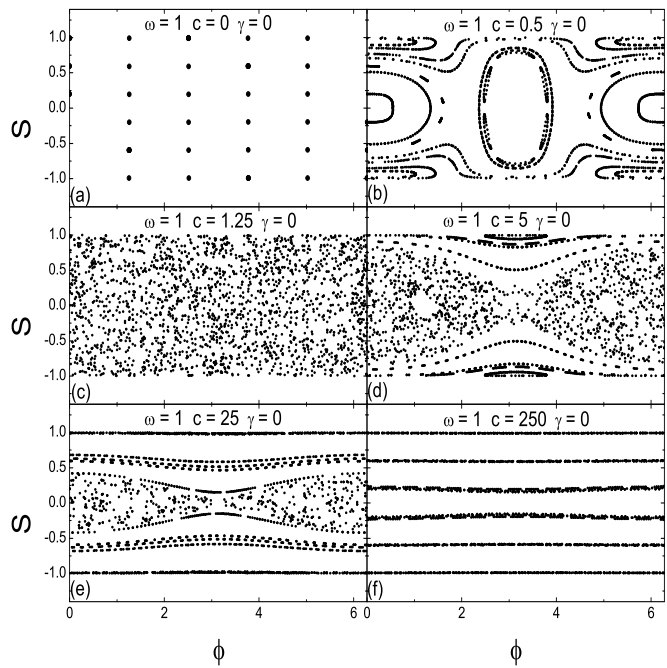}}
\end{center}
\par
\vspace*{-0.5cm} \caption{Phase space structure associated with the
classical effective Hamiltonian in Eq. (14) with $\beta=0$, in terms
of $s$ and $\phi$ defined in the text, for various values of $c$.
The driving frequency is given by $\omega=1$. As mentioned in the
text, all variables are appropriately scaled and hence
dimensionless.} \label{fig2}
\end{figure}

In Fig. 9 we plot $|b(\infty)|^{2}$ as a function of $\beta$, with
the values of $\omega$ same as in Fig. 8.  Consider first Fig. 9(a)
for the $c=0$ case. The $\beta$-dependence is seen to be rather
weak, in contrast to the $\cos^{2}(\beta)$ result in the RWA regime.
In addition, $|b(\infty)|^{2}$ is small for the entire range of
$\beta$. Next,  in Fig. 9(b) for $c=0.5$, the $\beta$ dependence
remains weak, but the final transition probability $|b(\infty)|^{2}$
becomes very significant in general: it can even reach almost 100\%
for certain values of $\beta$.  We hence observe
nonlinearity-enhanced population transfer again, but now outside the
RWA regime.  As $c$ is increased to $c=1.25$ in Fig. 9(c),  drastic
and full-range oscillations of $|b(\infty)|^{2}$ are observed.
Because Fig. 8(c) shows that this case corresponds to full
Hamiltonian chaos, the full-range oscillation in Fig. 9(c) is
clearly related to the underlying complete chaos on the mean-field
level.  As the nonlinearity parameter $c$ further increases, the
oscillation range of $|b(\infty)|^{2}$ decreases, consistent with
the observation from Fig. 8 that the chaotic layer becomes smaller
and smaller.  In the case of $c=75$ in Fig. 9(f), there is
essentially no transition probability for any value of $\beta$.  We
have also done more detailed analysis of how mean-field trajectories
should move as $\gamma$ is ramped, confirming that the size of the
chaotic layer determines the oscillation range of $|b(\infty)|^{2}$.

\begin{figure}[t]
\begin{center}
\vspace*{-0.5cm}
\par
\resizebox *{8.8cm}{8.8cm}{\includegraphics*{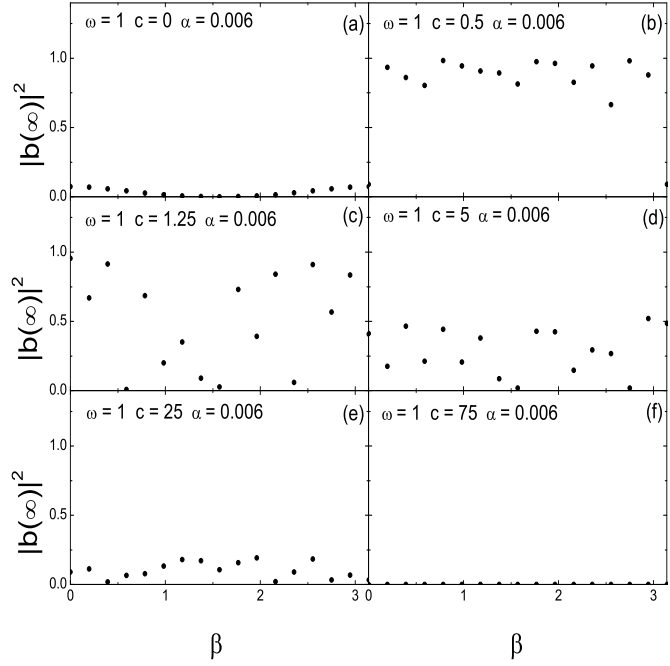}}
\end{center}
\par
\vspace*{-0.5cm} \caption{The final probability $|b(\infty)|^{2}$ as
a function of phase $\beta$ for $\alpha=0.006$ and $\omega=1$.
Results here should be connected with those in Fig. 8.} \label{fig2}
\end{figure}

Comparing results in Fig. 9(c) and 9(d) with those in Fig. 9(a) and
9(b), one finds another interesting aspect of driven NLZ dynamics.
That is, contrary to our observation in the RWA regime, nonlinearity
may also enhance the $\beta$-dependence of $|b(\infty)|^{2}$.  Note,
however, that the strong $\beta$-dependence observed here is due to
the mean-field chaos. Because chaos brings about strong
instabilities and hence large fluctuations in the quantum system,
results directly associated with the mean-field chaos should be
applied with caution. For example, the validity of the mean-field
treatment itself in the chaotic cases might be questionable when the
number of particles in a BEC is not extremely large.  This being so,
a low-frequency driving field generating the mean-field chaos could
be used to study effects of quantum fluctuations via the
$\beta$-dependence of $|b(\infty)|^{2}$.


\section {Conclusion}
Nonlinear extensions of LZ processes, and in particular, nonlinear
LZ processes in a periodic driving field, are of importance in
realizing the control of interacting many-body systems.  By
considering different parameter regimes of a periodic driving field,
we have exposed a number of interesting dynamical phenomena in
driven NLZ processes.  Many of these dynamical phenomena can be used
for the control of NLZ processes by a driving field.

In the high-frequency and large-amplitude regime, a driving field
can effectively generate new self-interaction terms that are absent
in non-driven two-mode NLZ models. The field can hence induce new
energy-level structures and can be useful in simulating new systems
not considered before. A driving-field-based scenario for the
suppression of undesired NLZ tunneling probabilities also becomes
possible.

In the high-frequency and small-amplitude regime,  driven NLZ
processes are also much different from driven LZ processes in at
least three aspects. First, for small nonlinearity strength the
final transition probability shows a shifted phase dependence on the
driving field. Second, for strong nonlinearity strength the
transition probability shows a much weakened phase dependence upon
the driving field. Third, in the adiabatic limit, significant
transition probabilities become a purely nonlinear effect, because
they were precisely zero in the absence of the self-interaction.
These differences between driven NLZ and driven LZ processes suggest
that a driving field may be used to directly expose and measure
nonlinear effects. Furthermore,  the adiabatic population transfer
based entirely on the nonlinear nature of the system can be almost
complete for certain amplitudes of the driving field,  thus leading
to an interesting approach to the improvement of adiabatic
population transfer in the presence of nonlinear Landau-Zener
tunneling.

In the third low-frequency regime, nonlinearity may also enhance the
final transition probability as well as its dependence on the phase
of the driving field. The enhanced phase dependence occurs for
intermediate nonlinear strength, where the NLZ dynamics displays
Hamiltonian chaos. A driving field in this regime can therefore be
used for studies of quantum fluctuations beyond the mean-field
level, as well as studies of manifestations of mean-field chaos in
the final transition probabilities.

To conclude, we have exposed many interesting aspects of driven
nonlinear Landau-Zener processes, by considering different parameter
regimes of a single-frequency and constant-amplitude driving field.
It is plausible that the driven nonlinear Landau-Zener processes are
even much richer in the presence of more complex driving fields,
such as those with multiple frequency components that have
time-evolving amplitudes and phases.  It should be also interesting
to extend our nonlinear Landau-Zener treatments to other many-body
generalizations of Landau-Zener processes that can go beyond the
mean-field level \cite{kayali,dob,altland}.

\acknowledgements One of the authors (J.G.) is supported by the
start-up funding (WBS grant No. R-144-050-193-101 and No.
R-144-050-193-133) and the NUS ``YIA'' funding (WBS grant No.
R-144-000-195-123), both from the National University of Singapore.
One of the authors (P.H.) acknowledges the financial support by the
Deutsche Forschungsgemeinschaft via the Collaborative Research
Centre SFB-486, project A10, and by the German Excellence Cluster
{\it Nanosystems Initiative Munich} (NIM) .

\bibliography{/home/bwu/references/berry}

\begin{thebibliography}{99}
\bibitem{Landau} Landau L D 1932 Phys. Z. Sowjetunion {\bf 2} 46

   Zener G 1932 Proc. R. Soc. London Ser. A {\bf 137} 696
\bibitem{LZ1} Ankerhold J and Grabert H 2003 Phys. Rev. Lett. {\bf 91} 016803

\bibitem{LZ12}Ithier G, Collin E, Joyez P, Vion D, Esteve D, Ankerhold J, and
Grabert H 2005 Phys. Rev. Lett. {\bf 94} 057004
\bibitem{LZ13} Zagoskin A M, Ashhab S, Johansson J R, and Nori F 2006
Phys. Rev. Lett. {\bf 97} 077001
\bibitem{LZ4} Wubs M, Kohler S, and H\"{a}nggi P 2007 Physica E {\bf 40}
187
\bibitem{LZ5}F\"{o}ldi P, Benedict M G, and Peeters F M 2008
Phys. Rev. A {\bf 77} 013406
\bibitem{LZ2} Saito K and Kayanuma Y 2004 Phys. Rev. B {\bf 70} 201304(R)
\bibitem{LZ3} Miyashita S 1995 J. Phys. Soc. Japan {\bf 64} 3207

 Wernsdorfer W and Sessoli R 1999 Science {\bf 284} 133
\bibitem{Hanggirev} Grifoni M and H\"{a}nggi P 1998 Phys. Rep. {\bf
304} 229
\bibitem{saitorev}Saito K, Wubs M, Kohler S, Kayanuma Y, and
H\"{a}nggi P 2007 Phys. Rev. B {\bf 75} 214308
\bibitem{Hanggi} Wubs M, Saito K, Kohler S, Kayanuma Y, and H\"{a}nggi P 2005  New J. Phys. {\bf 7} 218
\bibitem{holthaus}Holthaus M 2001 \pra{\bf 64} 011601
\bibitem{Liu} Liu J, Fu L, Ou B, Chen S, Choi D, Wu B, and Niu Q 2002 Phys. Rev. A {\bf 66} 023404
\bibitem{Wu} Wu B and Niu Q 2000 Phys. Rev. A {\bf 61} 023402
\bibitem{Korsch} Witthaut D, Graefe E M, and Korsch H J 2006 Phys. Rev. A {\bf 73}
063609

\bibitem{ZHG} Zhang Q, H\"anggi P, and Gong J B 2008 Phys.
Rev. A {\bf 77} 053607
\bibitem{saito08}Kayanuma Y and Saito K 2008 Phys. Rev. A {\bf 77}
010101(R)
\bibitem{modulationLiu} Wang G F, Fu L B, and Liu J 2006 Phys. Rev. A {\bf 73}
013619
\bibitem{CDT1991}
Grossmann F,  Dittrich T, Jung P, and H\"anggi P 1991 Phys. Rev.
Lett. {\bf 67} 516

\bibitem{devalle}
Della Valle G, Ornigotti M, Cianci E, Foglietti V, Laporta P, and
Longhi S 2007  Phys. Rev. Lett. {\bf 98} 263601

\bibitem{oberthaler}
Kierig E, Schnorrberger U, Schietinger A, Tomkovic J, and Oberthaler
M K  2008 Phys. Rev. Lett. {\bf 100}  190405


\bibitem{weiss} Weiss C and Teichmann N 2008 \prl{\bf 100} 140408
\bibitem{ober}Gati R and Oberthaler M K 2007 J. Phys. B {\bf 40} R61
\bibitem{double-well} Shin Y {\it et al} 2004 Phys. Rev. Lett. {\bf 92} 050405
\bibitem{oberprl}Albiez M, Gati R, F\"{o}lling J, Hunsmann S,
Cristiani M, and Oberthaler M 2005 \prl{\bf 85} 010402
\bibitem{two-band} Jona-Lasinio M {\it et al} 2003 Phys. Rev. Lett. {\bf 91} 230406
\bibitem{Rb}Matthews M R {\it et al} 1999 Phys. Rev. Lett. {\bf 83} 3358
\bibitem{luo}
Luo X, Xie Q, and Wu B 2007 \pra{\bf 76} 051802
\bibitem{kayali}Kayali M A and Sinitsyn N A 2003 Phys. Rev. A {\bf 67} 045603
\bibitem{dob} Dobrescu B E and Pokrovsky V L 2006 Phys. Lett. A {\bf
350} 154
\bibitem{altland} Altland A and Gurarie V 2008 Phys. Rev. Lett. {\bf 100} 063602


















\end{thebibliography}
\bibliographystyle{apsrev}

\end{document}